\documentstyle[epsfig,aps,floats]{revtex}

\begin{document}

\twocolumn[
\hsize\textwidth\columnwidth\hsize\csname@twocolumnfalse\endcsname

\title{Numerical Evidence for Continuity of
Mean Field and Finite Dimensional Spin Glasses}
\draft
\author{Enzo Marinari\cite{EMAIL}}
\address{Dipartimento di Fisica and INFN, 
Universit\`a di Cagliari,\\
via Ospedale 72, 09100 Cagliari (Italy)}

\date{\today}
\maketitle
\begin{abstract}
We study numerically a disordered model that interpolates among the 
Sherrington-Kirkpatrick mean field model and the three dimensional 
Edwards-Anderson spin glass.  We find that averages over the disorder 
of powers of the overlap and of the full $P(q)$ are smooth, and do not 
show any discontinuity.  Different lattice sizes are used to provide 
evidence for a smooth behavior of disorder averages in the 
thermodynamic limit.  Quantities defined on a given realization of the 
disorder show a chaotic behavior.  Our results support the validity of 
a Replica Symmetry Breaking description of finite dimensional models.
\end{abstract}


\twocolumn
\vskip.5pc ] 
\narrowtext


The relation among the mean field spin glass model \cite{SK} (SK) and 
its finite dimensional, realistic counterpart \cite{EA} (EA), is a 
debated and interesting subject.  If on one side the Parisi replica 
symmetry breaking (RSB) solution \cite{PARISI,MEPAVI} of the mean 
field theory is getting a stronger acceptance even on a rigorous 
ground \cite{GUERRA,AIZCON,PARSUM}, on the other side the relation of 
RSB theory with finite dimensional models is less clear.

Numerical simulations (see \cite{MAPARU} and references therein) seem 
to confirm the validity of the Parisi picture in $3$ and $4$ spatial 
dimensions $D$, but the droplet model approach \cite{DROPLET} suggests 
a different behavior of finite dimensional models.  Recent analytical, 
rigorous and heuristic work can be seen as hinting for potential 
troubles in the application of RSB to finite dimensional models 
\cite{NEWSTE}, or as confirming the suitability of such a description 
\cite{NONEWS,MPRRZ}.

If the droplet model description \cite{DROPLET} applies to finite
dimensional spin glasses the mean field model (that we assume as
described by the Parisi RSB Ansatz) has to be a singular case: at
variance with the usual Ising model the case of $D=\infty$ would be
different in nature from the case of any finite dimensionality. All
the exceptional features that appear in the description of the mean
field model solution would disappear for any large but finite
dimension $D$. Typical equilibrium configurations of the SK and the EA
models would have to be very different. On the contrary validity of 
the RSB ansatz in the finite dimensional case would imply that 
phenomena like the existence of a phase transition in non-zero 
magnetic field, and some form of ultrametric 
organization of the states appear also for finite dimensional models.

Here we analyze numerically this issue, and show that there are no
discontinuities in the typical behavior of the system (clearly the
behavior of a single system turns to be, as expected, very
discontinuous, while average over the disorder, i.e. the quenched
state, does not show any abrupt discontinuity). There are many contexts
where this result can be relevant. For example Guerra \cite{GUERRA} has
shown rigorously the validity of relations like \cite{MEPAVI} 

\begin{equation}
E( q^2_{1,2} q^2_{2,3} ) 
= \frac12 E( q^4_{1,2}) +
  \frac12 E( q^2_{1,2})^2\ ,
\protect\label{E-GUERRAA}
\end{equation}
and

\begin{equation}
E( q^2_{1,2} q^2_{3,4} )
= \frac13 E( q^4_{1,2}) +
  \frac23 E( q^2_{1,2})^2\ ,
\protect\label{E-GUERRAB}
\end{equation}
where $q_{a,b}$ if the overlap $\frac{1}{N} \sum_i\sigma_i^a 
\sigma_i^b$ among replica $a$ and replica $b$ of the system, and 
$E(\cdot)$ denotes the expectation value over the thermal noise and 
over the quenched disorder.  The proof holds for mean field, and, 
under a continuity hypothesis, for finite dimensional models.  
Numerically one finds \cite{MPRRZ} that in $3D$ relations like 
(\ref{E-GUERRAA},\ref{E-GUERRAB}) are satisfied at better than one 
part over thousand (with $E(q^2)$ different from zero in the spin 
glass phase).  This is one of the reasons for which checking the 
continuity of the limit of finite dimensionality is an important task.

Another relevant example where a weak perturbation is used to derive 
important relations among equilibrium observables and dynamic 
quantities that can be measured in experiments is reference 
\cite{FMPP}.  Here we will analyze a system where the perturbation 
does not need to be small: we want to understand the nature of the 
relation among the mean field model SK and the finite dimensional EA 
Ising spin glass.  In our case starting from a mean field interaction 
involving with $p>2$ spin would not give a smooth limit to a two spin 
interaction three dimensional model, since the mean field starting 
point would be a one step RSB $P(q)$.

We have defined and studied a model that interpolates from mean field
to $3D$.  The Hamiltonian of our model is a function of the parameter
$\epsilon$, which ranges from zero to one:

\begin{equation}
  H_\epsilon \equiv (1-\epsilon) H_{3D} + \epsilon H_{SK}\ .
\end{equation}
All the couplings $J_{i,j}$ are binary, and take with probability one
half the values $\pm 1$. The three dimensional structure of the system
is given by the term

\begin{equation}
  H_{3D} \equiv \sum_{\langle i,j \rangle} \sigma_i J^{(1)}_{i,j} \sigma_j\ ,
\end{equation}
where the sum runs over first neighboring sites on a simple cubic
lattice in three dimensions, with $N=L^3$ sites.
The mean-field, Sherrington Kirkpatrick like contribution is
given by the term

\begin{equation}
  H_{SK} \equiv \frac{1}{N} \sum_{i,j} \sigma_i J^{(2)}_{i,j} \sigma_j\ ,
\end{equation}
where the sum runs over all site couples of the lattice.  The quenched 
couplings $\{ J^{(1)}, J^{(2)} \}$ are independent.  For $\epsilon=0$ 
we recover the $3D$ EA spin glass, while for $\epsilon=1$ we recover 
the SK mean field model.

We have focused our analysis on the probability distribution of the 
overlap, $P^\epsilon(q)$, as a function of $\epsilon$, averaged over 
the quenched disorder, and on the $P^\epsilon_J(q)$ for fixed 
realization of the disorder $\{ J^{(1)}, J^{(2)} \}$.  We have also 
analyzed in detail the second, fourth and sixth momentum $q^2$, $q^4$ 
and $q^6$.

As we will discuss our results clearly show that $P^\epsilon(q)$ is a
smooth function of $\epsilon$ in all the range of $\epsilon$ ranging
from zero to one, for, say $T\simeq 0.7 T_c$.  On the contrary for a
given disorder sample the shape of $P_J^\epsilon(q)$ depends
dramatically on $\epsilon$. The expectation values of the first even
powers of the overlap all have a smooth dependence over
$\epsilon$. Larger lattices show a behavior consistent with this
picture, and we do not see any singular behavior developing. In the
clear limits of a numerical simulation we can state are observing a
smooth transition from the mean field to the finite dimensional model.

The {\em Parallel Tempering} Monte Carlo approach (for a review see
for example \cite{TEMPER}) has been crucial for thermalizing our
systems down to $T\simeq 0.7 T_c$. Since $T_c=1$ for the mean field
model and $T_c\simeq 1.11$ \cite{KAWYOU} is a reasonable estimate for
the $3D$ system with binary couplings we have analyzed our data as a
function of $\epsilon$ at fixed $T=0.7$. Renormalizing the $T$ value
as a function of $\epsilon$ could have been in principle a more exact
procedure: we have decided not to do so not to add a further degree of
freedom (the scaling of the temperature) that would have in itself
potentially added a chaotic behavior, without adding much to the
interpretation of the result. We are looking for smoothness of the
observables (on the state averaged over the disorder) and not for
precise extrapolations, so we can afford ignoring the small
renormalization of $T_c$ (that we could only control approximately in
any case). All the data we will present in the following will be at
$T=0.7$. 

This numerical simulation is expensive, since for all $\epsilon$ 
values we pay the price ($N^2$) of a mean field simulation (this is 
only the scaling of the number of operations needed, and does not 
include an estimate for the increment of the needed computer time due 
to critical slowing down, so that it has to be read as a lower bound 
to the increase of the computer time): the time taken from a lattice 
sweep scales like the volume squared instead than like the volume.

At $L=8$ we have been able to analyze $150$ disorder samples, each 
with $11$ $\epsilon$ values (from $0$ to $1.0$ with steps of $0.1$), 
while for $L=12$ we have $11$ disorder samples with the same 
$\epsilon$ values.  As we have said this is not a small computational 
effort because of the factor volume added by the mean field 
interaction term and the factor eleven for the different $\epsilon$ 
values that multiply the total computer time.  We have used a 
multispin coded program (for the long range part of the interaction).  
The simulations have taken a few years of a nowaday typical 
workstation.

The Parallel Tempering Monte Carlo updating scheme worked effectively 
at all our $\epsilon$ values for both $L=8$ and $L=12$.  For $L=8$ we 
have used $13$ copies of the system, with temperatures starting with 
$0.7$ and increasing with steps of $0.05$ up to $1.3$, while at $L=12$ 
we have used $26$ copies with the same minimum value of $T$, up to 
$T=1.95$ with the same step.  We ran $100.000$ lattice sweeps (and 
tempering updates) for each disorder realization and $\epsilon$ value 
both at $L=8$ and at $L=12$.  At all values of the parameters we have 
discarded the first half of the Monte Carlo sweeps for sake of 
thermalization.  At each new $\epsilon$ value we restarted the system 
from a random configuration, even when running the same disorder 
sample (in order to avoid any bias toward continuity).  All 
thermalization tests \cite{TEMPER} were satisfactory: the $T$ swap 
acceptance ratio was included among $0.3$ and $0.7$, and each copy of 
the system had eventually spent time in all allowed $T$ values.

We have used the two different lattice sizes ($L=8$ and $L=12$) to get
control over the finite size scaling of our results. It turns out that
all the features we have revealed and we will discuss now are stable
when increasing the lattice size, and make us confident we are
observing an asymptotic behavior. The $L=12$ results have a larger
error than the $L=8$ ones (since in this case the $N^2$ scaling of the
computer time needed for the simulation has limited us to $11$
samples), but they confirm the behavior we are observing at $L=8$.

\begin{figure}[p]
\begin{center}
\leavevmode
\centering\epsfig{file=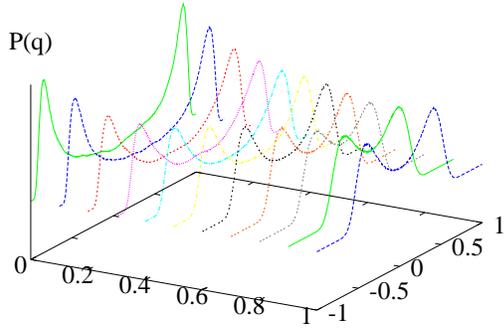,angle=270,width=1.0\linewidth}
\end{center}
\caption[0]{$P^{\epsilon}(q)$ as a function of $\epsilon$ (going from 
$0$ to $1$) and $q$ (from $-1$ to $+1$). $L=8$, $T=0.7$,
disorder average taken over  $150$ realizations.}
\label{FIG-PQ}
\end{figure}

In figure (\ref{FIG-PQ}) we show the average probability distribution 
of the overlap $q$, $P(q)$, for the $L=8$ lattice, as a function of 
$\epsilon$ (and $q$), at $T=0.7$. The $\epsilon$ axis goes 
from zero to one, the overlap axis from $-1$ to $1$.  The behavior of 
the function $P^{\epsilon}(q)$ as a function of $\epsilon$ is very 
smooth.  It is important to notice that the value of $P(q=0)$ at 
$\epsilon=0$ is largely different from zero (in a definitely 
significant way: this is very well known, see for example 
\cite{MAPARU} and references therein).  Continuity of 
$P^{\epsilon}(q)$ as a function of $\epsilon$ seems very clear from 
figure (\ref{FIG-PQ}).  For $L=12$ we get a very similar picture (see 
also the value of $q^{2}(\epsilon)$ that we will show later on), and 
things stay as smooth as they are in figure (\ref{FIG-PQ}).
The non trivial probability distribution that one finds in the mean 
field approach survives smoothly down to the finite dimensional, 
realistic system.

\begin{figure}[p]
\begin{center}
\leavevmode
\centering\epsfig{file=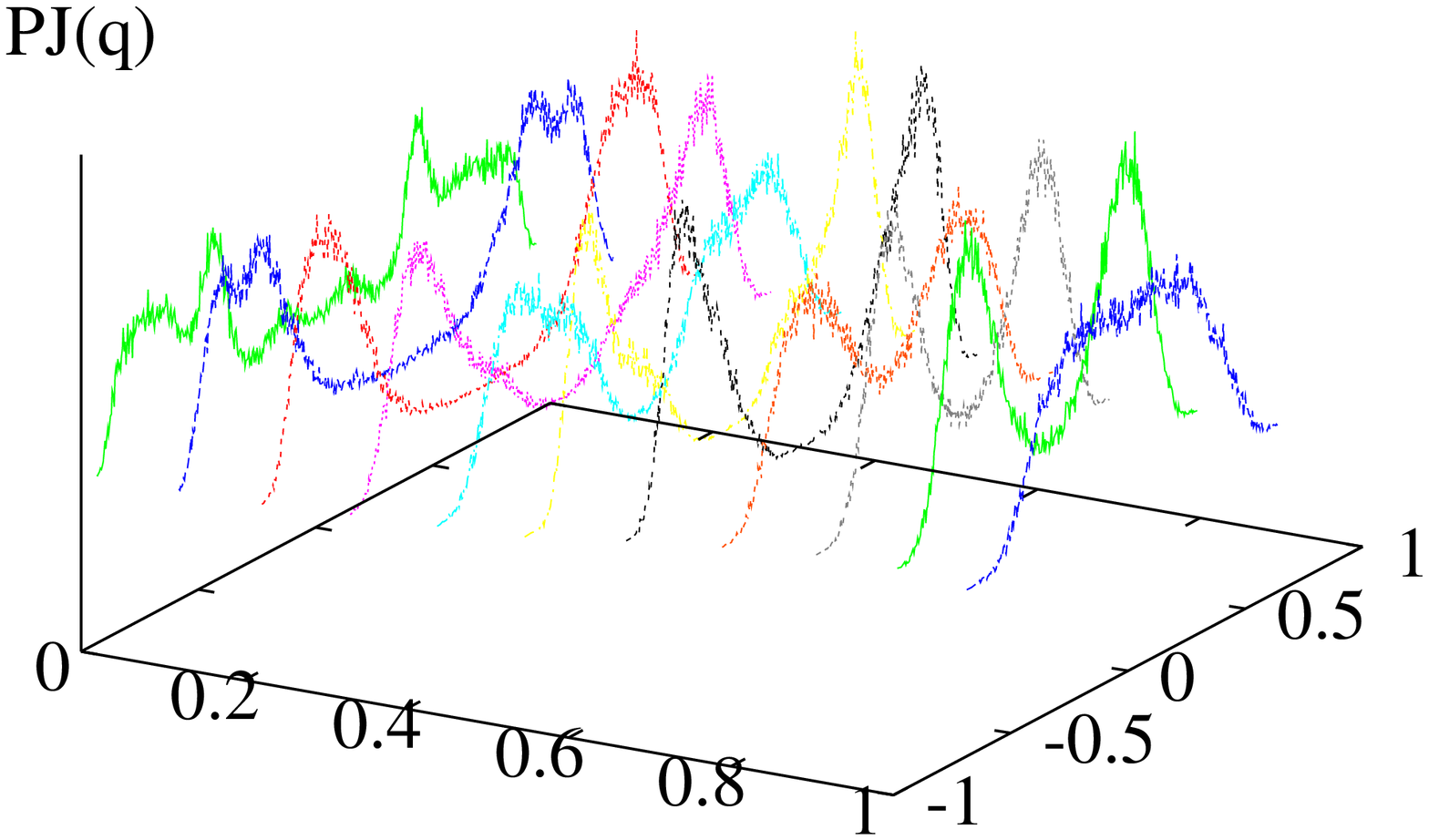,angle=270,width=1.0\linewidth}
\end{center}
\caption[0]{As in figure (\ref{FIG-PQ}), but $P_{J}^{\epsilon}(q)$
for one realization of 
the quenched disorder.} 
\label{FIG-PQJA}
\end{figure}

\begin{figure}[p]
\begin{center}
\leavevmode
\centering\epsfig{file=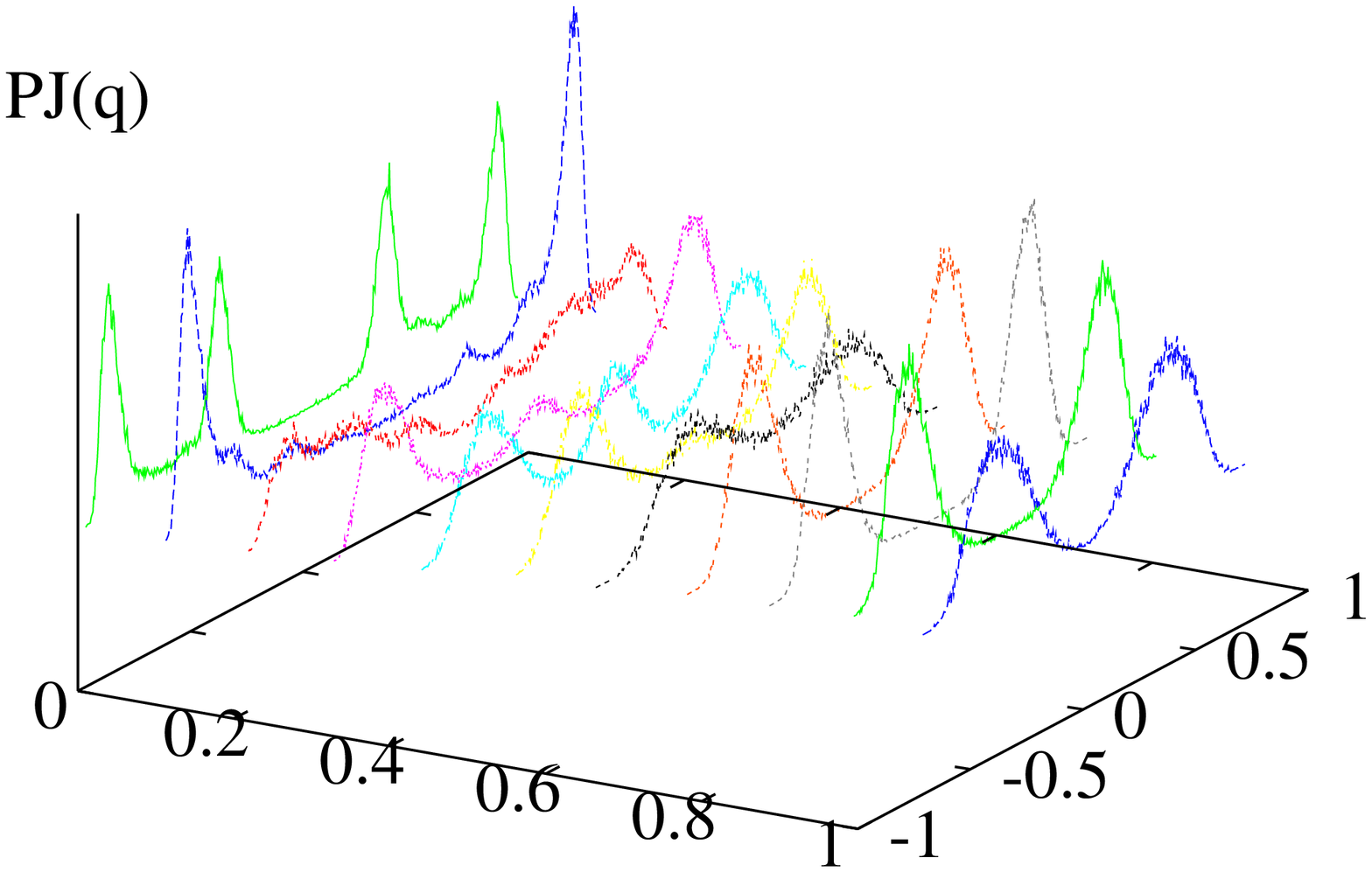,angle=270,width=1.0\linewidth}
\end{center}
\caption[0]{As in figure (\ref{FIG-PQ}), but $P_{J}^{\epsilon}(q)$
for a second realization of 
the quenched disorder.} 
\label{FIG-PQJB}
\end{figure}

We do not expect continuity for a given realization of the quenched 
disorder: the mean field picture would suggest that the typical 
equilibrium configurations are highly unstable under variations of 
$\epsilon$ \cite{MEPAVI}.  Our results confirm this point of view in a 
very nice way.  We show in figures (\ref{FIG-PQJA}) and 
(\ref{FIG-PQJB}) two $P_{J}^{\epsilon}(q)$ for two typical 
realizations of the quenched disorder.  In both figures one sees that 
the shape of $P_{J}^{\epsilon}(q)$ varies completely with $\epsilon$.  
In figure (\ref{FIG-PQJA}) the mean field model is in a configuration 
that carries all the weight close to $q=0$.  Already at $\epsilon=0.9$ 
the system is completely different, and $P_{J}^{0.9}(q)$ has two 
maxima.  Notice that the level of symmetry of these 
$P_{J}^{\epsilon}(q)$ is a good check of the level of thermalization 
of our Monte Carlo tempered runs.  The pure $3D$ system has in this 
case a broad distribution, with two clear, small maxima, and weight 
also at $q\simeq 0$.

In the sample of figure (\ref{FIG-PQJB}) instead the $\epsilon=1$ mean 
field model shows two broad maxima.  Again the evolution is chaotic, 
down to the $\epsilon=0$ $3D$ model where there are four sharp maxima. We 
want to notice that, if any, the structure of the individual  
$P_{J}^{\epsilon}(q)$ becomes more complex when approaching the pure 
$3D$ model: structures with many sharp peaks appear on the larger 
lattices more in the $3D$ model than in the mean field one. One 
reason for that is probably connected to the fact that finite size 
effects are larger in the mean field model. Recently a detailed 
discussion of models that show a ``modified droplet model'' structure 
has been reported in \cite{BOVIER}. This is very interesting since 
provides a rigorous construction of models with a complex structure of 
states, but looks different from what we find here. $P_{J}^{\epsilon}(q)$  
like the one we observe at $\epsilon=0$ in figure (\ref{FIG-PQJB}) 
would not appear in a ``modified droplet model'' structure: our 
results hint that we are dealing with a complete RSB like structure of 
states.
 
\begin{figure}[p]
\begin{center}
\leavevmode
\centering\epsfig{file=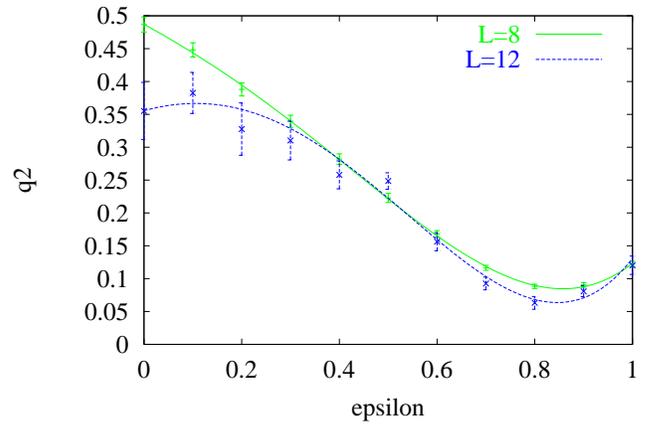,angle=270,width=1.0\linewidth}
\end{center}
\caption[0]{$E(q^{2})$ as a function of 
$\epsilon$ for $L=8$ and $L=12$.} 
\label{FIG-Q2}
\end{figure}

In figure (\ref{FIG-Q2}) we show $E(q^{2})$ as a function of 
$\epsilon$ for $L=8$ and $L=12$.  The two lines are the best fit of 
the data to a third order polynomial.  The most part of this (smooth) 
dependence over $\epsilon$ can be related to the renormalization of 
$T$.  The $L=12$ data show an even smaller dependence over $\epsilon$ 
than the $L=8$ data.  We have also analyzed the fourth and the sixth 
momentum of the overlap $q$, that have the same smooth behavior.

We acknowledge valuable conversations with Giorgio Parisi.


\end{document}